\documentstyle[12pt]{article}

\newcommand{\sect}[1]{\setcounter{equation}{0}\section{#1}}

\newcommand{\EQ}{\begin{equation}}
\newcommand{\EN}{\end{equation}}
\newcommand{\bea}{\begin{eqnarray}}
\newcommand{\ena}{\end{eqnarray}}
\newcommand{\vs}[1]{\vspace{#1 mm}}
\newcommand{\hs}[1]{\hspace{#1 mm}}
\renewcommand{\a}{\alpha}
\renewcommand{\b}{\beta}
\renewcommand{\c}{\gamma}
\renewcommand{\d}{\delta}
\newcommand{\e}{\epsilon}
\def\bbox{{\,\lower0.9pt\vbox{\hrule \hbox{\vrule height 0.2 cm
\hskip 0.2 cm \vrule height 0.2 cm}\hrule}\,}}
\newcommand{\dsl}{\pa \kern-0.5em /}
\newcommand{\la}{\lambda}
\newcommand{\shalf}{\frac{1}{2}}
\newcommand{\pa}{\partial}

\newcommand{\td}{{\tilde d}}

\newcommand{\nn}{\nonumber\\}

\begin{document}

\topmargin 0pt
\oddsidemargin 5mm

\begin{titlepage}
\setcounter{page}{0}
\begin{flushright}
OU-HET 256 \\
hep-th/9701095
\end{flushright}

\vs{10}
\begin{center}
{\Large NON-EXTREME BLACK HOLES FROM INTERSECTING M-BRANES}

\vs{15}
{\large
Nobuyoshi Ohta\footnote{e-mail address: ohta@phys.wani.osaka-u.ac.jp}
and
Takashi Shimizu\footnote{e-mail address: simtak@phys.wani.osaka-u.ac.jp}}\\
\vs{10}
{\em Department of Physics, Osaka University, \\
Toyonaka, Osaka 560, Japan}

\end{center}
\vs{15}
\centerline{{\bf{Abstract}}}
\vs{5}

We investigate the possibility of extending non-extreme black hole solutions
made of intersecting M-branes to those with two non-extreme deformation
parameters, similar to Reissner-Nordstr{\o}m solutions. General analysis of
possible solutions is carried out to reduce the problem of solving field
equations to a simple algebraic one for static spherically-symmetric case
in $D$ dimensions. The results are used to show that the extension to
two-parameter solutions is possible for $D=4,5$ dimensions but not for
higher dimensions, and that the area of horizon always vanishes in
the extreme limit for black hole solutions for $D \geq 6$ except for two
very special cases which are identified. Various solutions are also
summarized.

\end{titlepage}
\newpage
\renewcommand{\thefootnote}{\arabic{footnote}}
\setcounter{footnote}{0}

\sect{Introduction}

There has been much interest in solitonic and black hole solutions in
string theories because of the possible resolution of various puzzles
associated with quantum gravity~\cite{DGH,HS,GU,Sen,CYe,CYn,CT1}.
The study of this subject has got its upsurge due to the recent discovery
that the statistical origin of the entropy of BPS-saturated black holes
can be addressed from the string point of
view~\cite{SV,CM,BMPV,MSS,MS,JKM,HS1,HMS,HLM}.

By now many extreme as well as non-extreme solutions have been understood
from the low-dimensional effective theory.
There is now a consensus that the best candidate for a unified theory
underlying all physical phenomena is no longer ten-dimensional string theory
but rather eleven-dimensional M-theory. Though the precise formulation of
M-theory is not known, its low energy limit coincides with
the eleven-dimensional supergravity. Thus it is expected that the black hole
solutions get their natural framework in the eleven-dimensional supergravity.

Indeed, the most transparent and systematic approach to this problem is
provided by identifying these black holes as compactified configurations of
intersecting two-branes and five-branes of eleven-dimensional
supergravity~\cite{PT}. It has been known that this theory admits various
$p$-brane solutions~\cite{GU}, collectively referred to as M-branes.
These solutions have been shown to be understood as the intersections of
the fundamental two- and five-brane
solutions~\cite{PT,T,DLP,PO,KT1,KT2,CT,GA,DR}.

Now rather systematic (the so-called ``harmonic function'') rule is
developed to produce various extreme solutions as intersecting M-branes in
eleven dimensions~\cite{T}, and most of the known solutions can be derived
systematically upon dimensional reduction to lower dimensions. This rule
has also been extended to the one for non-extreme solutions with a single
``non-extremality'' parameter specifying a deviation from the
BPS-limit~\cite{CT}. It would be quite interesting to further elaborate on
this approach and investigate how large a class of solutions are allowed.

In this paper we examine the possibility of extending non-extreme black
hole solutions composed of intersecting M-branes in eleven dimensions to
those with two deformation parameters, similar to the Reissner-Nordstr{\o}m
one in four dimensions. We perform a general analysis of possible solutions
for the static spherically-symmetric case in $D$ dimensions and summarize
various typical solutions as well as new ones. Upon dimensional reduction,
these eleven-dimensional solutions give rise to various non-extreme solutions
in lower dimensions. In the extreme limit, most of these solutions give
supersymmetric ones~\cite{CH}.\footnote{It turns out that these solutions
in $D=4,5$ dimensions are equivalent to known solutions~\cite{GU,CT} by
redefinitions of variables and parameters to be discussed shortly.}
However, our analysis is not restricted to supersymmetric case and should
be useful to searching for new solutions.

There has been a claim that there are no extreme black holes with
regular horizons of finite area in $D \geq 6$~\cite{CYn,Sen,KT1,CT}.
Though no counter example to this claim has been discovered, general
proof allowing the possibility of two non-extremality parameters has not
been given even for the static case of our interest. This is also one of
our motivations for trying to find solutions with two deformation
parameters, since then we can keep a deformation parameter finite to
obtain a horizon of finite area in the extreme limit. We use the results
of the general analysis of the solutions to prove that the above claim
is indeed true except only two solutions in $D=6,7$ dimensions.
However, it can be shown that supersymmetry is not recovered in the
``extreme'' limit in the latter special solutions. If we require that there
should be exact supersymmetry in the limit, our results indicate that there
are no black holes with horizons of finite area in higher dimensions.
We also discuss the universal expressions for the ADM mass and entropy of
these black hole solutions.

The paper is organized as follows. In sect.~2, we summarize elements of
eleven-dimensional supergravity relevant to our study, and collect various
formulae used in the following sections. We then present our non-extreme
black hole solutions in the form with two deformation parameters as
intersecting M-branes for $D=4$ in sect.~3 and for $D=5$ in sect.~4.
The general analysis of possible solutions for $D \geq 6$ dimensions is
presented in sect.~5, where we also show that there are no solutions with
two deformation parameters and that almost no extreme black holes can
have regular horizons of finite area. In sect.~6, we summarize typical
higher-dimensional solutions, including the six- and seven-dimensional
solutions with regular horizons of finite area in the extreme limit.
In sect.~7, we discuss the universal expressions for ADM mass and entropy.
Sect.~8 is devoted to conclusions.

\sect{Intersecting M-Branes}

In this section, we start with various definitions and conventions used
in this paper. Our basic strategy is to look for the solutions in the
low-energy limit of the M-theory, eleven-dimensional supergravity.

The bosonic parts of the field equations for eleven-dimensional supergravity
are
\bea
R_{MN} &=& \frac{1}{3} (F^2)_{MN} - \frac{1}{36} F^2 g_{MN},\nn
\nabla_M F^{MNPQ} &=& - \frac{1}{576} \e^{NPQM_1 \cdots M_8}
 F_{M_1 \cdots M_4} F_{M_5 \cdots M_8}, \nn
\pa_{[M} F_{NPQR]} &=& 0.
\label{fe}
\ena
where $M, N, \cdots = 0,1, \cdots ,10$ are the curved-space indices and
$\nabla_M$ is the covariant derivative. The last eq. in (\ref{fe}) is the
Bianchi identity. It is also useful to note that
$\nabla_M F^{MNPQ} = \pa_M (\sqrt{-g} F^{MNPQ})/\sqrt{-g}$ where $g$
is the determinant of the metric.

The vanishing condition of the supersymmetry transformation of the
gravitino gives the criterion for the existence of the unbroken supersymmetry.
This is called the  Killing equation for the Killing spinor $\zeta$:
\EQ
\left[ \pa_M + \frac{1}{4} \omega_M^{ab} \Gamma_{ab}
 + \frac{1}{144} \left( {\Gamma_M}^{NPQR}- 8 \d_M^N\Gamma^{PQR}
 \right)F_{NPQR}\right]\zeta = 0.
\label{ki}
\EN
where $\omega_M^{ab}$ is the spin connection, $a,b$ are the tangent-space
indices and $\Gamma$'s are the antisymmetrized products of eleven-dimensional
gamma matrices with unit strength. We are going to look for solutions for
the field equations~(\ref{fe}), and examine how supersymmetry remains
unbroken by using the Killing equation~(\ref{ki}).

We take the following metric for our system:
\bea
ds_{11}^2 = -e^{2u_0} dt^2 + e^{2u_1} \widehat{dy_1}^2
 + \sum_{\a=2}^{d-1} e^{2 u_\a} dy_\a^2 + e^{2v} dr^2
 + e^{2B} r^2 d\Omega_{\td+1}^2,
\label{metric}
\ena
where the coordinate $(t,y_\a), (\a=1,\ldots, d-1)$ parametrize the
$d$-dimensional world-volumes of the intersecting M-branes with
\EQ
\widehat{dy_1} = dy_1 + e^w dt.
\label{off}
\EN
The remaining coordinates of the eleven-dimensional spacetime are the radius
$r$ and the angular coordinates on a $(\td+1)=(10-d)$-dimensional unit sphere,
whose metric is $d\Omega_{\td+1}^2$. The off-diagonal component of the metric
(\ref{off}) is introduced in order to allow for the possibility of
incorporating momentum along the M-brane space direction ($y_1$).
All the functions appearing in the metrics are assumed to depend only on
the radius $r$ of the transverse dimensions, which, together with the time
$t$, will be our eventual spacetime of dimension $D=\td+3$.

Typically these metrics are chosen to be the products of various harmonic
functions. The rule for identifying which coordinates belong to which
brane is as follows~\cite{T}. If one divides out from the eleven-dimensional
metric the overall conformal factor which multiplies the transverse
spacetime, the coordinates multiplied with the same harmonic function belong
to a given $p$-brane. We will soon see how this rule works.

In order to solve for our eqs.~(\ref{fe}), we need the Ricci tensor for
our metric~(\ref{metric}). It is straightforward to derive the following
results:
\bea
R_{00} &=& e^{2(u_0-v)} \left[ u_0'' + u_0' \left\{ u_0' + \sum_{\a=1}^{d-1}
 u_\a' - v' + (\td + 1) \left( B'+\frac{1}{r} \right) \right\} \right] \nn
&& - e^{2(u_1 + w - v)} \left[ u_1'' + w'' + u_0' (u_1'- w') + 2 u_1' w'
 + \frac{3}{2}(w')^2 \right. \nn
&& \left. \hs{25} + (u_1' + w') \left\{ \sum_{\a=1}^{d-1} u_\a' -v' + (\td+1)
 \left( B' + \frac{1}{r} \right) \right\} \right] \nn
&& -\frac{(w')^2}{2} e^{2(- u_0 + 2 u_1 - v + 2w)}, \nn
R_{01} &=& - e^{2(u_1 - v) + w} \left[ u_1'' + \frac{w''}{2}
 + u_0' \left( u_1' - \frac{w'}{2} \right)  \right. \nn
&& \left. \hs{25} + \left( u_1' + \frac{w'}{2} \right) \left\{
 \sum_{\a=1}^{d-1} u_\a' + w' - v' + (\td+1) \left( B' + \frac{1}{r} \right)
 \right\} \right] \nn
&& -\frac{(w')^2}{2} e^{2(-u_0 + 2 u_1 - v) + 3w}, \nn
R_{11} &=& - e^{2(u_1 - v)} \left[ u_1'' + u_1' \left\{ u_0'
 + \sum_{\a=1}^{d-1} u_\a' - v' + (\td+1) \left( B' + \frac{1}{r} \right)
 \right\} \right] \nn
&& -\frac{(w')^2}{2} e^{2(-u_0 + 2 u_1 - v + w)}, \nn
R_{\a\b} &=& - e^{2(u_\a - v)} \left[ u_\a'' + u_\a' \left\{ u_0'
 + \sum_{\c=1}^{d-1} u_\c' - v' + (\td+1) \left( B'
 + \frac{1}{r} \right) \right\} \right] \d_{\a\b}, \;\; (\a,\b>1) \nn
R_{rr} &=& - u_0'' - u_0'( u_0'- v') - \sum_{\a=1}^{d-1} \left[ u_\a'' +
 u_\a' (u_\a' - v') \right]  \nn
&& - (\td+1) \left[ B'' + \frac{2B' - v'}{r} - v'B' + (B')^2 \right]
 + \frac{(w')^2}{2} e^{2(-u_0 + u_1 + w)}, \nn
R_{ab} &=& - e^{2(B-v)} \left[ B'' + \left( B' + \frac{1}{r} \right)
 \left\{ u_0' + \sum_{\a=1}^{d-1} u_\a' - v' + (\td +1) \left( B'
 + \frac{1}{r} \right) \right\} - \frac{1}{r^2} \right] g_{ab} \nn
&& \hs{15} + \frac{\td}{r^2} g_{ab},
\label{ricci}
\ena
where a prime denotes a derivative with respect to $r$, and $g_{ab}$ is
the metric for $(\td+1)$-sphere of radius $r$.

These formulae are valid not only for eleven-dimensional case but also
for other dimensions. In particular, these will be useful for examining
solutions for string theories in ten dimensions as well as the theories
compactified in lower dimensions.

\sect{$D=4$ non-extreme solutions}

We first present our non-extreme solutions with two deformation parameters
which correspond to $D=4$ black holes. In this section $\td=1$.
Four-dimensional extreme black holes with four charges can be derived from
the two different intersecting M-brane configurations toroidally
compactified~\cite{T,KT2}: two two-branes and two five-branes
($2\perp 2\perp 5\perp 5$) and three five-branes with a boost along the
common string ($5\perp 5\perp 5$). Although our non-extreme solutions prove
to be equivalent to known ones~\cite{GU,CT}, it is instructive to see the
explicit forms and their relations. This also serves to establish our
notation. We will be very brief.

\subsection{Non-extreme intersecting $2\perp 2\perp 5\perp 5$ M-branes}

The first of the above eleven-dimensional configurations corresponds to
the intersecting $2\perp 2\perp 5\perp 5$. Our metric is given by
\bea
ds_{11}^2 &=& (H_1 H_2)^{1/3}(H_3 H_4)^{2/3} \left[ - (H_1 H_2 H_3 H_4)^{-1}
 f dt^2 + H_1^{-1} (H_3^{-1}dy_1^2 + H_4^{-1}dy_2^2) \right. \nn
&& + H_2^{-1} (H_3^{-1} dy_3^2 + H_4^{-1} dy_4^2)
 + (H_3 H_4)^{-1} (dy_5^2 + dy_6^2 +dy_7^2) \nn
&& \left. + f^{-1} dr^2 + r^2 d\Omega_2^2 \right],
\label{2255}
\ena
which is a non-extreme generalization of the extreme solutions discussed
in ref.~\cite{KT2} and the same form taken in ref.~\cite{CT}.

Our non-extreme deformation is parametrized by the two-parameter function
\EQ
f = \left(1- \frac{\mu_1}{r^\td} \right) \left( 1 - \frac{\mu_2}{r^\td}
 \right).
\label{nondf}
\EN
The coordinates $y_1, \ldots, y_7$ will describe the toroidally compactified
directions.

As explained in sect.~2, up to an overall factor, (the inverse of) the
harmonic function $H_1$ multiplies the squares of the differentials of the
coordinates $t,y_1$ and $y_2$, which means that
these coordinates belong to a two-brane. Examining
how other harmonic functions multiply the coordinates, one can easily see
that this metric corresponds to a configuration of non-extreme intersecting
$2\perp 2\perp 5\perp 5$ M-branes.

We assume the following forms for our metric~(\ref{2255}):\footnote{Throughout
this paper, we use a prime to denote a derivative with respect to $r$, and
never use it to distinguish different functions.}
\bea
F_{012r} &=& \shalf {T_1}'; \;\;
F_{034r} = \shalf {T_2}'; \;\;
F_{24ab} = \shalf\e_{abr} {T_3}'; \;\;
F_{13ab} = - \shalf\e_{abr} {T_4}', \nn
H_i &=& 1 + \frac{Q_i}{r}, \;\; (i=1,2,3,4),
\label{non1}
\ena
where the indices $a$ and $b$ to the field strengths denote the angular
coordinates of the transverse dimensions.

We substitute these into the field eqs.~(\ref{fe}) and look for the solution
for the field strength. We find the solution
\bea
T_i &=& 1 - \frac{\sqrt{(Q_i + \mu_1)(Q_i + \mu_2)}}{r + Q_i},\;\;
 {\rm for} \;\; i=1,2, \nn
T_j &=& 1 - \frac{\sqrt{(Q_j + \mu_1)(Q_j + \mu_2)}}{r},\;\;
 {\rm for} \;\; j=3,4.
\label{sol2255}
\ena

The non-extreme solutions discussed in ref.~\cite{CT} correspond to $\mu_2=0$.
Alternatively, if we make the redefinition
\EQ
{\hat r}^\td \equiv r^\td - \mu_2; \hs{3}
{\hat Q}_i \equiv Q_i + \mu_2; \hs{3}
{\hat \mu} \equiv \mu_1 -\mu_2,
\label{repl}
\EN
with $\td=1$, our solution reduces to that in ref.~\cite{CT}.\footnote{
This was pointed out to us by R. Emparan and by I. Klebanov and A. Tseytlin.}
In that case, the outer and inner horizons sit at ${\hat r}={\hat\mu^{1/\td}}$
and 0, and the extreme limit is given by ${\hat \mu} \to 0$. Thus the
deformation parameter $\mu_2$ plays the role of shifting the horizon from
$r=0$ to $\mu_2^{1/\td}$. An advantage of our expression is that it makes
manifest that the area of the outer horizon is finite in the extreme limit.

Suppose $\mu_1>\mu_2$. It is convenient to parametrize
\EQ
Q_i = - \frac{\mu_1+\mu_2}{2} + \frac{\mu_1-\mu_2}{2} \cosh 2\d_i.
\label{mu1}
\EN
The nine-area at the outer horizon $r=\mu_1$ is then given by
\bea
A_9 &=& 4 \pi L^7 \left[ r^2 (H_1 H_2 H_3 H_4)^{1/2} \right]_{r=\mu_1}
 = 4 \pi L^7 \prod_{i=1}^4 \sqrt{Q_i + \mu_1} \nn
&=& 4 \pi L^7 (\mu_1-\mu_2)^2 \prod_{i=1}^4\cosh \d_i,
\ena
where $L$ is the (common) length of the coordinates $y_\a$.

The extreme limit is obtained by sending $\mu_1,\mu_2 \to \mu$ while keeping
$(\mu_1-\mu_2)^{1/2} \cosh \d_i$ (or $Q_i$) finite. In what sense is this
an ``extreme limit''? Of course, the outer and inner horizons coincide in
this limit. Moreover, examining the condition for the existence of non-zero
Killing spinors~(\ref{ki}), one finds that indeed $1/8$ supersymmetry is
preserved in the limit even for non-zero $\mu$. This is the same number
of remaining supersymmetry for the extreme solution with $f=1$.
In this limit, the nine-area becomes
\EQ
(A_9)_{BPS} = 4 \pi L^7 \sqrt{(Q_1+\mu)(Q_2+\mu)(Q_3+\mu)(Q_4+\mu)}.
\EN

If we compactify $y_1, \cdots ,y_7$, our solution reduces to the
four-dimensional black hole with the Einstein-frame metric
\EQ
ds_4^2 = - \la(r) f(r)dt^2 +\la^{-1}(r)\left[ f^{-1}(r)dr^2+r^2 d\Omega_2^2
\right],
\label{4met}
\EN
\EQ
\la(r) = (H_1 H_2 H_3 H_4)^{-1/2}
= \frac{ r^2}{
\left[ (r+Q_1)(r+Q_2)(r+Q_3)(r+Q_4) \right]^{1/2}}.
\label{la1}
\EN
Under the replacement~(\ref{repl}), the four-dimensional metric
(\ref{4met}) reduces to the solutions with two electric and two magnetic
charges found in ref.~\cite{CYn,CT}.

Dimensional reduction to $D=10$ along one direction common to the two
five-branes, say $y_7$, gives type IIA solution representing R-R $p$-brane
configuration $2\perp 2\perp 4\perp 4$. T-duality along one of the two
directions common to four-branes transforms this to the $3\perp 3\perp 3
\perp 3$ solution of type IIB. Reduction along $y_1$, on the other hand,
produces the $1\perp 2\perp 4\perp 5$ solution. Other solutions are
similarly obtained by using T- and $SL(2,Z)$ duality.

\subsection{Non-extreme intersecting $5\perp 5\perp 5$ M-branes with boost}

The non-extreme generalization of the extreme $5\perp 5\perp 5$ M-branes
with boost is given by
\bea
ds_{11}^2 &=& (H_1 H_2 H_3)^{2/3} \left[ (H_1 H_2 H_3)^{-1} (- K^{-1} f dt^2
 + K \widehat{dy_1}^2 ) + (H_2 H_3)^{-1} (dy_2^2 + dy_3^2) \right. \nn
&& + (H_1 H_3)^{-1} (dy_4^2 + dy_5^2) + (H_1 H_2)^{-1} (dy_6^2 + dy_7^2)
 \left. + f^{-1} dr^2 + r^2 d\Omega_2^2 \right],
\ena
where
\bea
\widehat{dy_1} &=& dy_1 + \frac{\tilde K(r)}{K(r)} dt,\hs{3}
F_{23ab} = \shalf \e_{abr} {T_1}';\hs{3}
F_{45ab} = \shalf \e_{abr} {T_2}';\hs{3}
F_{67ab} = \shalf \e_{abr} {T_3}'; \nn
K(r) &=& 1 + \frac{P}{r}; \;\;
H_i(r) = 1 + \frac{Q_i}{r}, \;\; (i=1,2,3).
\ena
Here $a$ and $b$ are again the coordinates of the transverse dimensions other
than $r$, and $f$ is the same as eq.~(\ref{nondf}) with $\td=1$, and the
charge $P$ stands for the momentum along the common string in $y_1$ direction.

The solution to the field eqs.~(\ref{fe}) is given by
\bea
{\tilde K} = -\frac{\sqrt{(P+\mu_1)(P+\mu_2)}}{r}; \hs{3}
T_i = 1 - \frac{\sqrt{(Q_i + \mu_1)(Q_i + \mu_2)}}{r}, \;\;
(i=1,2,3).
\ena
Like (\ref{mu1}), we parametrize the charges by
\bea
Q_i = - \frac{\mu_1+\mu_2}{2} + \frac{\mu_1-\mu_2}{2} \cosh 2\d_i;\hs{3}
P = - \frac{\mu_1+\mu_2}{2} + \frac{\mu_1-\mu_2}{2} \cosh 2\c.
\label{mu2}
\ena
We again find the nine-area at the outer horizon $r=\mu_1$ to be
\bea
A_9 &=& 4 \pi L^7 \left[ r^2 (K H_1 H_2 H_3)^{1/2} \right]_{r=\mu_1}
 = 4 \pi L^7 \sqrt{(Q_1 + \mu_1)(Q_2 + \mu_1)(Q_3 + \mu_1)(P + \mu_1)} \nn
&=& 4 \pi L^7 (\mu_1-\mu_2)^2 \cosh \d_1 \cosh \d_2 \cosh \d_3 \cosh \c.
\ena
In the extreme limit $\mu_1,\mu_2 \to \mu$ keeping $(\mu_1-\mu_2)^{1/2}
\cosh \d_i (\c)$ or $Q_i$ and $P$ finite, the nine-area becomes
\EQ
(A_9)_{BPS} = 4 \pi L^7 \sqrt{(Q_1+\mu)(Q_2+\mu)(Q_3+\mu)(P+\mu)}.
\EN

The corresponding four-dimensional Einstein-frame metric is (\ref{4met}) with
\EQ
\la(r) = (H_1 H_2 H_3 K)^{-1/2} = \frac{r^2}
{\left[ (r+Q_1)(r+Q_2)(r+Q_3)(r+P)\right]^{1/2}},
\label{la2}
\EN
in agreement with the metric found in ref.~\cite{CYn,CT} (under the
replacement~(\ref{repl})) with one electric and three magnetic charges.

Compactification to $D=10$ along the direction $y_1$ common to three
five-branes gives rise to type IIA solutions corresponding to three
four-branes intersecting over two-branes plus additional Kaluza-Klein
electric charge background, which may be called $0\perp 4\perp 4\perp 4$
solution. T-duality transforms this solution to $2\perp 2\perp 4\perp 4$
which is the dimensional reduction of the $2\perp 2\perp 5\perp 5$
solution mentioned in the previous subsection. Upon compactification in
other direction, say $y_2$, gives type IIA $5\perp 4\perp 4$ solution with
boost. Again other solutions may be obtained by T- and $SL(2,Z)$ duality.

\sect{$D=5$ non-extreme solutions}

The extreme $D=5$ black holes with three charges can be obtained from
the two different configurations~\cite{T}: three two-branes intersecting
at a point ($2\perp 2\perp 2$) and intersecting two-brane and five-brane
along a common string ($2\perp 5$) with boost. We now briefly summarize
the non-extreme solutions which correspond to $D=5$ black holes.

\subsection{Non-extreme intersecting $2\perp 2\perp 2$ M-branes}

For the intersecting $2\perp 2\perp 2$ M-branes, our metric is given by
\bea
ds_{11}^2 &=& (H_1 H_2 H_3)^{1/3} \left[ - (H_1 H_2 H_3)^{-1}f dt^2
 + H_1^{-1} (dy_1^2 + dy_2^2) + H_2^{-1} (dy_3^2 + dy_4^2) \right. \nn
&& \left.+ H_3^{-1} (dy_5^2 + dy_6^2) + f^{-1} dr^2 + r^2 d\Omega_3^2 \right].
\label{222}
\ena
In this section, our non-extreme deformation is parametrized by~(\ref{nondf})
with $\td=2$.

We take our metrics and field strengths of the following forms:
\bea
H_i &=& 1 + \frac{Q_i}{r^2}, \;\; (i=1,2,3), \nn
F_{012r} &=& \shalf {T_1}'; \;\;
F_{034r} = \shalf {T_2}'; \;\;
F_{056r} = \shalf {T_3}'.
\label{an222}
\ena
Substituting these into the field eqs.~(\ref{fe}), we again find that
there is a solution
\EQ
T_i = 1 - \frac{\sqrt{(Q_i + \mu_1)(Q_i + \mu_2)}}{r^2 + Q_i}, \;\;
(i=1,2,3).
\label{sol222}
\EN
For $Q_1=Q_2=Q_3$ and $\mu_2=0$, this solution becomes the anisotropic
six-brane solution in ref.~\cite{GU}. For $\mu_2=0$ or under the
replacement~(\ref{repl}), this coincides with the solution in ref.~\cite{CT}.

If we use the parametrization (\ref{mu1}),
the nine-area at the outer horizon $r=\sqrt{\mu_1}$ is given by
\bea
A_9 &=& 2 \pi^2 L^6 \left[ r^3 (H_1 H_2 H_3)^{1/2} \right]_{r=\mu_1^{1/2}}
 = 2 \pi^2 L^6 \prod_{i=1}^3 \sqrt{Q_i + \mu_1} \nn
&=& 2 \pi^2 L^6 (\mu_1-\mu_2)^{3/2} \prod_{i=1}^3 \cosh \d_i.
\ena

In the extreme limit $\mu_1,\mu_2 \to \mu$ while keeping $(\mu_1-\mu_2)^{1/2}
\cosh \d_i$ finite, supersymmetry is recovered and the area becomes
\EQ
(A_9)_{BPS} = 2 \pi^2 L^6 \sqrt{(Q_1+\mu)(Q_2+\mu)(Q_3+\mu)}.
\EN

Upon toroidal compactification of $y_1, \cdots ,y_6$, our solution reduces
to the five-dimensional black hole with the Einstein-frame metric
\EQ
ds_5^2 = - \la^2(r)f(r)dt^2 +\la^{-1}(r)\left[ f^{-1}(r)dr^2+r^2 d\Omega_3^2
\right],
\label{5met}
\EN
\EQ
\la(r) = (H_1 H_2 H_3)^{-1/3}
 = \frac{r^2}{ \left[(r^2+Q_1)(r^2+Q_2)(r^2+Q_3) \right]^{1/3}}.
\label{la3}
\EN
Under the replacement~(\ref{repl}), this agrees with the metric of
non-extreme five-dimensional black holes found in ref.~\cite{CYn,HMS,CT},
with three electric charges.

\subsection{Non-extreme intersecting $2\perp 5$ M-branes with boost}

The non-extreme generalization of the supersymmetric configuration of a
two-brane intersecting with a five-brane boosted along the common
string~\cite{T} takes the form
\bea
ds_{11}^2 &=& H_1^{1/3} H_2^{2/3} \left[ (H_1 H_2)^{-1} (- K^{-1} f dt^2
 + K \widehat{dy_1}^2 ) + H_1^{-1} dy_2^2 \right. \nn
&& \left. + H_2^{-1} (dy_3^2 + \cdots + dy_6^2) + f^{-1} dr^2
 + r^2 d\Omega_3^2 \right],
\label{25}
\ena
where
\bea
\widehat{dy_1} &=& dy_1 + \frac{\tilde K(r)}{K(r)} dt; \hs{3}
F_{012r} = \shalf {T_1}'; \hs{3}
F_{2abc} = \shalf \e_{abcr} {T_2}', \nn
K(r) &=& 1 + \frac{P}{r^2}; \hs{3}
H_i(r) = 1 + \frac{Q_i}{r^2}, (i=1,2).
\label{an25}
\ena
Here $a,b,c$ denote the angular coordinates of the transverse dimensions,
and $f$ is the same as eq.~(\ref{nondf}). The charge $P$ corresponds to
the momentum along the direction $y_1$.

Plugging these into the field eqs.~(\ref{fe}), we find the solution
\bea
{\tilde K} &=& - \frac{\sqrt{(P + \mu_1)(P + \mu_2)}}{r^2},\nn
T_1 &=& 1 - \frac{\sqrt{(Q_1 + \mu_1)(Q_1 + \mu_2)}}{r^2 + Q_1}; \hs{3}
T_2 = 1 - \frac{\sqrt{(Q_2 + \mu_1)(Q_2 + \mu_2)}}{r^2}.
\ena

If we use the parametrization~(\ref{mu2}),
we again find that the nine-area at the outer horizon is given by
\bea
A_9 &=& 2 \pi^2 L^6 \left[ r^3 (K H_1 H_2)^{1/2} \right]_{r=\mu_1^{1/2}}
 = 2 \pi^2 L^6 \sqrt{(Q_1+\mu_1)(Q_2+\mu_1)(P+\mu_1)} \nn
&=& 2 \pi^2 L^6 (\mu_1-\mu_2)^{3/2} \cosh \d_1 \cosh \d_2 \cosh \c.
\ena

The extreme limit is again to send $\mu_1,\mu_2 \to \mu$ while keeping
$(\mu_1-\mu_2)^{1/2} \cosh \d_i (\c)$ or $Q_i$ and $P$ finite.
In this limit, the nine-area becomes
\EQ
(A_9)_{BPS} = 2 \pi^2 L^6 \sqrt{(Q_1+\mu)(Q_2+\mu)(P+\mu)}.
\EN

The corresponding five-dimensional Einstein-frame metric is (\ref{5met}) with
\EQ
\la(r) = (H_1 H_2 K)^{-1/3}
 = \frac{r^2}{\left[ (r^2+Q_1)(r^2+Q_2)(r^2+P)\right]^{1/3}},
\label{la4}
\EN
basically the metric found in ref.~\cite{CYn,HMS,CT}, with two electric and
one magnetic charges.

\sect{General analysis of $D\geq 6$ non-extreme solutions}

We have examined if the same method applies to higher dimensional case
such as two-branes. One of our motivations for this analysis is that
if we can have nonzero $\mu_2$, we can have extreme solutions with
nonzero horizon area since then we need not set the non-extremality
parameters to zero but to equal finite value.

It is certainly true that if we put $H_2=H_3=1$
in our intersecting $2\perp 2\perp 2$ M-branes~(\ref{222}), this gives
a special solution corresponding to a two-brane solution in eleven-dimensions,
with additional backgrounds of field strengths $F_{034r}$ and $F_{056r}$.
However, it turns out that these cannot be extended to higher dimensions
unless $\mu_2=0$. Below we present a general analysis of the
higher-dimensional solutions and show that the problem of solving field
equations boils down to a simple algebraic algorithm.

We use this result to show that there is no extreme solutions with regular
horizons of finite area except for special ones which are identified.
It should be noted that our following analysis is not restrited to
supersymmetric case in the extreme limit.

\subsection{Metrics}

Let us start the general analysis of solutions in higher dimensions.
In order to see if there is any solution, it is sufficient to examine the
case of equal charges. It is straightforward to generalize the results
to solutions with different charges afterwards, as we will show in the next
section. Hence we take the following metric as the most general static
spherically-symmetric one (boost charge will be considered later):
\bea
ds_{11}^2 = - H^{2a_0}f dt^2 + \sum_{\a=1}^{d-1} H^{2a_\a} dy_\a^2
 + H^{2b} \left( f^{-1} dr^2 + r^2 d\Omega_{\td+1}^2 \right),
\ena
where $H=1+\frac{Q}{r^\td}$ is a harmonic function in $(\td+2)$ dimensions
with the charge $Q$, $f$ is given in eq.~(\ref{nondf}), and $a$'s and $b$
are constants to be determined.

A straightforward calculation using (\ref{ricci}) yields
\bea
R_{00} &=& H^{2(a_0-b)} f \left[
 - a_0 \frac{\td^2 {\tilde Q}^2}{H^2 r^{2\td +2}}
 + (a_0+1) \frac{\td^2 \mu_1\mu_2}{r^{2\td +2}}\right], \nn
R_{\a\b} &=& H^{2(a_\a-b)} \left[
 a_\a \frac{\td^2 {\tilde Q}^2}{H^2 r^{2\td +2}}
 - a_\a \frac{\td^2 \mu_1\mu_2}{r^{2\td +2}}\right] \d_{\a\b},
 \hs{3} (\a,\b=1,\ldots,d-1), \nn
R_{rr} &=& \frac{1}{f} \left[
 a_0 \frac{\td^2 {\tilde Q}^2}{H^2 r^{2\td +2}}
 - \left(\sum_{\a=0}^{d-1} a_\a^2 + a_0 + \td b^2 -b \right)
 f \frac{\td^2 Q^2}{H^2 r^{2\td +2}}
 - (a_0+1) \frac{\td^2 \mu_1\mu_2}{r^{2\td +2}}\right], \nn
R_{ab} &=& \left[ b \frac{\td^2 {\tilde Q}^2}{H^2 r^{2\td +2}}
 - \td (\td b -1 ) \frac{\mu_1\mu_2}{r^{2\td +2}}\right] g_{ab},
\label{ricci1}
\ena
where we have defined ${\tilde Q}^2 = (Q+\mu_1)(Q+\mu_2)$.
In deriving the result~(\ref{ricci1}), we have imposed the condition
\EQ
\sum_{\a=0}^{d-1} a_\a + \td b = 0.
\label{cond1}
\EN
Unless this condition is obeyed, there remain quite complicated terms which
cannot satisfy field equations. Also this is satisfied by all known solutions.

From the field eq.~(\ref{fe}), we have
\EQ
(F^2)_{MN} = 3R_{MN} + 3 R g_{MN},
\label{mn}
\EN
which, together with (\ref{ricci1}), yields
\bea
(F^2)_{ab} &=& \left[ 3 (a_0+2b) \frac{\td^2 {\tilde Q}^2}{H^2 r^{2\td +2}}
 - 3 \left(\sum_{\a=0}^{d-1} a_\a^2 + a_0 + \td b^2 - b \right)
 f \frac{\td^2 Q^2}{H^2 r^{2\td +2}} \right. \nn
&& \left. - 3 \td \left\{ \td(a_0+2b+1)-2 \right\}
 \frac{\mu_1\mu_2}{r^{2\td +2}} \right] g_{ab}.
\label{ab}
\ena

Up to this point, we have made no assumption on dimensionality. From
now on, we specialize to $D \geq 6 (\td \geq 3)$. There are two cases
to be discussed separately.

\subsection{The case of vanishing $(F^2)_{ab}$}

Let us first consider the case in which $(F^2)_{ab}$ vanishes. This is
always the case for $D\geq 7(\td \geq 4)$ since we have only fourth-rank
antisymmetric tensor. From the first two terms in eq.~(\ref{ab}),
we must have
\EQ
b=- \frac{1}{2} a_0; \hs{3}
\sum_{\a=1}^{d-1} a_\a^2 = - \frac{\td+4}{4} a_0^2 - \frac{3}{2} a_0,
\label{cond2}
\EN
which gives
\EQ
(F^2)_{ab} = - 3 \td(\td-2) \frac{\mu_1\mu_2}{r^{2\td +2}} g_{ab}.
\EN
This does not vanish for $\td \geq 3$ unless $\mu_2=0$, implying that
(\ref{cond2}) is valid and $\mu_2$ must vanish.

With the condition (\ref{cond2}) and $\mu_2=0$, we find from (\ref{ricci1})
and (\ref{mn})
\bea
(F^2)_{00} &=& - H^{3a_0} f \frac{9}{2} a_0
 \frac{\td^2 {\tilde Q}^2}{H^2 r^{2\td+2}}, \nn
(F^2)_{\a\a} &=& H^{(2 a_\a+a_0)} \frac{3}{2}(2 a_\a + a_0)
 \frac{\td^2 {\tilde Q}^2}{H^2 r^{2\td+2}}, \nn
(F^2)_{rr} &=& \frac{1}{f} \frac{9}{2} a_0
 \frac{\td^2 {\tilde Q}^2}{H^2 r^{2\td+2}}.
\label{back2}
\ena

So far, we have made no assumption on the background field
strengths. In order to satisfy the field equations, let us introduce all
possible terms for the field strengths\footnote{We are not considering the
case in which KK monopoles exist~\cite{CO}. It is easy to see that this
does not affect our main results.}
\EQ
F_{0\a\b r} = \shalf a_{\a\b} {T_1}',
\label{back1}
\EN
where $a_{\a\b}$ are constants antisymmetric in their indices.
Equating the contribution from (\ref{back1}) to (\ref{back2}), we obtain
\bea
{T_1}' = H^{a_0 + a_\a + a_\b - 1} \frac{\td {\tilde Q}}{r^{\td+1}},
\label{sol1}
\ena
and
\bea
\sum_{\a<\b} a_{\a\b}^2 &=& - 3 a_0, \nn
\sum_{\b (\neq \a)} a_{\a\b}^2 &=& - (2 a_\a + a_0 ),
 \hs{3} {\rm for} \hs{3} \a=1,\ldots, d-1.
\label{cond3}
\ena
Consistency with the field eq.~(\ref{fe}) requires that only the
coefficients $a_{\a\b}$ that satisfy
\EQ
a_0 + a_\a + a_\b = -1,
\label{cond4}
\EN
can be nonvanishing. In this case eq.~(\ref{sol1}) gives
\EQ
T_1 = 1 - \frac{\tilde Q}{r^\td + Q}.
\label{t1}
\EN
Thus our problem of solving field equations boils down to the simple
algebraic one of finding solutions to (\ref{cond1}), (\ref{cond2}) and
(\ref{cond3}). Simple solutions are
\bea
{\rm Two-brane}: && a_0= a_1= a_2= -\frac{1}{3}; \hs{2}
a_3 = \cdots = a_{d-1} = b =\frac{1}{6}; \hs{2}
a_{12}=1.
\label{2}
\\
2\perp 2:&& a_0=- \frac{2}{3}; \; a_1=a_2=a_3=a_4= -\frac{1}{6}; \;
a_5 = \cdots = a_{d-1} = b = \frac{1}{3}; \nn
&& a_{12}=a_{34}= 1,
 {\rm (others = 0)}.
\label{22}
\ena

If there are monopole backgrounds, the last condition~(\ref{cond3}) is
slightly changed, but the general properties of the solutions remain the same.

These results are enough to establish that there are no solutions with
regular horizons of finite area with just one exception.
From the condition (\ref{cond2}), one finds
\EQ
0 > a_0 > - \frac{6}{\td+4}.
\label{range1}
\EN
On the other hand, the nine-area is given by
\EQ
A_9 = \omega_{\td + 1} L^{d-1} \left[ r^{\td +1}
 H^{\sum_{\a=1}^5 a_\a + (\td +1) b} \right]_{r=\mu_1^{1/\td}},
\EN
which, with help of eqs.~(\ref{cond1}) and (\ref{cond2}), is cast into
\bea
A_9 &=& \omega_{\td + 1} L^{d-1} \left[ r^{\td +1} H^{-3a_0/2}
 \right]_{r=\mu_1^{1/\td}}, \nn
&\to& \omega_{\td + 1} L^{d-1} \mu_1^{(\td +1)/\td + 3a_0/2} Q^{-3a_0/2},
\label{nine1}
\ena
near the extreme limit $\mu_1 \sim 0$. Here $\omega_{\td+1}$ is the volume
of the unit $(\td+1)$-sphere
\EQ
\omega_{\td+1} = \frac{2\pi^{\frac{\td}{2}+1}}
 {\Gamma\left(\frac{\td}{2} + 1 \right)}.
\EN
However, the area (\ref{nine1}) vanishes for (\ref{range1}) in the
extreme limit.

If we can incorporate boost, it introduces the factor $K^{1/2}$ which
has the effect of reducing the exponent of $\mu_1$ in (\ref{nine1}) by 1/2.
This means that we have the possibility of nonvanishing nine-area
for $a_0=-(\td+2)/3 \td$. To have the boost, we must have a null isometry
in another direction, say $y_1$~\cite{GAR}. Thus we must have
\EQ
a_0=a_1= - \frac{\td+2}{3 \td}.
\label{ok}
\EN
Substituting (\ref{ok}) back into the second condition in (\ref{cond2})
shows that there is a unique solution for $\td=4$:
\bea
&& a_0 = a_1 = - \shalf; \hs{2}
a_2 = a_3 = a_4 = 0; \hs{2}
b = \frac{1}{4}; \nn
&& a_{1\a} = \frac{1}{\sqrt{2}}, (\a=2,3,4),
\label{stfi1}
\ena
where use has also been made of the conditions (\ref{cond3}) in deriving
$a_{1\a}$. The field strengths $F_{01\a r} (\a=2,3,4)$ are nonvanishing
and provide sources for 2-branes lying in the planes $(1,\a)$. We thus
see that this configuration corresponds to three 2-branes intersecting over
a string $(2^3)$. This is one of the two cases in which the nine-area does
not vanish in the extreme limit. (The other case will be given in the
next subsection.) For all other cases, the nine-area always vanishes in
the limit.\footnote{The solutions with monopoles discussed in
ref.~\cite{CO} suggest that there is no other case with nonzero area.}
This establishes the promised result.

\subsection{The case of nonvanishing $(F^2)_{ab}$}

Let us turn to the case of nonvanishing $(F^2)_{ab}$. This is possible
only for $\td=3$, since we can then introduce
\EQ
F_{abcd} = \shalf \e_{abcdr} {T_2}',
\EN
where the Bianchi identity in eq.~(\ref{fe}) requires that $T_2$ should be
a harmonic function. This gives
\EQ
(F^2)_{ab} = \frac{3}{2} ({T_2}')^2 H^{-6b} g_{ab}.
\label{fab}
\EN
Equating this with the first term in eq.~(\ref{ab}), we must have
\bea
b=\frac{1}{3}; \hs{3}
\sum_{\a=0}^{5} a_\a^2 = - a_0 - 3 b^2 + b ;\hs{3}
3(a_0+2b+1)=2,
\ena
in order to have nonvanishing $\mu_2$. However, this leads to
\EQ
(F^2)_{ab} = - 9 \frac{{\tilde Q}^2}{H^2 r^{8}} g_{ab},
\EN
which is negative and cannot be consistent with eq.~(\ref{fab}).
This means that we must have $\mu_2=0$. Similar reasoning excludes the
possibility of equating (\ref{fab}) with other terms in (\ref{ab}) unless
$\mu_2=0$. We thus conclude that $\mu_2$ must again be zero.

Putting $\mu_2=0$, we learn that either the first term or second term in
(\ref{ab}) must balance with the contribution from the field
strengths~(\ref{fab}). It is easy to exclude the second possibility by
an analysis similar to the above. So we are left with the first
possibility:
\EQ
b = \frac{1}{3}; \hs{3}
\sum_{\a=1}^{5} a_\a^2 = - a_0 (a_0 +1).
\label{cond5}
\EN
We then have the solution for this part
\EQ
T_2 = \sqrt{\frac{2}{3}(3 a_0 + 2)}\left(1- \frac{\tilde Q}{r^3}\right).
\label{solab}
\EN

With the condition (\ref{cond5}), we find from (\ref{ricci1}) and (\ref{mn})
\bea
(F^2)_{00} &=& - H^{2(a_0-1/3)} f (6 a_0+1)
 \frac{9 {\tilde Q}^2}{H^2 r^{8}}, \nn
(F^2)_{\a\a} &=& H^{2(a_\a-1/3)} (3 a_\a + 3 a_0 +1)
 \frac{9 {\tilde Q}^2}{H^2 r^{8}}, \nn
(F^2)_{rr} &=& \frac{1}{f} (6 a_0 +1)
 \frac{9 {\tilde Q}^2}{H^2 r^{8}}.
\ena

As our field strengths, we introduce (\ref{back1}). The same analysis as
above shows that $T_1$ is given by eq.~(\ref{t1}) with $\td=3$ and
\bea
\sum_{\a<\b} a_{\a\b}^2 &=& - \frac{2}{3}(6 a_0 +1), \nn
\sum_{\b (\neq \a)} a_{\a\b}^2 &=& - \frac{2}{3}(3 a_\a + 3 a_0 +1),
 \hs{3} {\rm for} \hs{3} \a=1,\ldots, 5,
\label{cond6}
\ena
where only those $a_{\a\b}$ are nonvanishing when (\ref{cond4}) is
satisfied.

Thus our problem again reduces to the algebraic one of finding solutions
to (\ref{cond1}), (\ref{cond4}), (\ref{cond5}) and (\ref{cond6}).
Simple examples of the solutions are
\EQ
{\rm Five-brane}:\hs{3} a_0= a_1= \cdots = a_5= -\frac{1}{6}; \hs{2}
a_{\a\b}=0.
\label{5}
\EN
\EQ
2 \subset 5: \hs{3} a_0= a_1=a_2=-\frac{1}{3}; \hs{2}
a_{12}=\sqrt{\frac{2}{3}}; \hs{2}
{\rm (others = 0)}.
\label{25i}
\EN
These examples with boost will be summarized in the next section.

Let us finally examine if we can have solutions with horizons of finite area
in the extreme limit. From the reality of (\ref{solab}) and (\ref{cond5}),
$a_0$ is restricted to be
\EQ
0 > a_0 > - \frac{2}{3}.
\label{range2}
\EN
On the other hand, with the use of eq.~(\ref{cond1}), the nine-area is
given by
\EQ
A_9 = \omega_4 L^5 \left[ r^4 H^{b-a_0} \right]_{r=\mu_1^{1/3}}
 \to \omega_4  L^5 \mu_1^{1+a_0} Q^{-a_0+1/3},
\label{nine2}
\EN
near the extreme limit. This vanishes for (\ref{range2}) in the extreme limit.

What about the possibility of introducing boost? Since this has the effect
of reducing the exponent in (\ref{nine2}) by 1/2, we have the
possibility of getting nonvanishing nine-area for $a_0=-\shalf$.
Choosing $y_1$ for the direction of the null isometry, we again find a unique
solution
\bea
&& a_0=a_1=-\shalf; \hs{2}
a_2=\cdots=a_5=0, \nn
&& a_{1\a} = \frac{1}{\sqrt{3}}, (\a=2,\ldots,5); \hs{2}
{\rm other \; }a{\rm 's} =0,
\label{stfi2}
\ena
where use has been made of the conditions (\ref{cond5}), (\ref{cond6}) and
(\ref{cond4}). This is similar to the solution~(\ref{stfi1}) and allows the
interpretation of four 2-branes and a 5-brane intersecting over a string
$(2^4\subset 5)$. These two solutions (\ref{stfi1}) and (\ref{stfi2})
constitute the only cases in which the nine-area does not vanish in the
extreme limit.

We thus conclude that $\mu_2=0$ necessarily in higher dimensions and
that the nine-area must vanish in the extreme limit except for the two
special cases of seven- and six-dimensional ones (\ref{stfi1}) and
(\ref{stfi2}).

\sect{$D\geq 6$ non-extreme solutions}

In this section, we summarize typical solutions for higher dimensions
for completeness, including the interesting case of the intersecting
two-branes with regular horizons of finite area. Though some of them are
known ones~\cite{GU,CT}, it is instructive to see how the solutions with
equal charges considered in the previous section are generalized and
also to see how these are consistent with the previous analysis.

In what follows, we take $\mu_2=0$ and
\bea
\widehat{dy_1} &=& dy_1 + \frac{\tilde K(r)}{K(r)} dt; \hs{2}
{\tilde Q}^2 = Q(Q+\mu_1); \hs{2}
{\tilde P}^2 = P(P+\mu_1), \nn
f(r) &=& 1-\frac{\mu_1}{r^\td}; \hs{2}
H(r) = 1 + \frac{Q}{r^\td}; \hs{2}
K(r) = 1 + \frac{P}{r^\td}; \hs{2}
{\tilde K} = -\frac{\tilde P}{r^\td}.
\label{nondf1}
\ena

\subsection{$2^3$ with boost in $D=7$}

The boosted solution (\ref{stfi1}) identified to have nonvanishing area
has the metric
\bea
ds_{11}^2 &=& H^{1/2} \left[ H^{-3/2} (- K^{-1} f dt^2 + K \widehat{dy_1}^2 )
 + H^{-1/2}( dy_2^2 + dy_3^2 + dy_4^2) \right. \nn
&& \left. + f^{-1} dr^2 + r^2 d\Omega_5^2 \right],
\label{14}
\ena
for $\td=4$. It does not seem possible to have different charges in this
case. The solution is
\bea
F_{01\a r} = \frac{1}{2 \sqrt{2}} {T_1}', (\a=2,3,4); \hs{3}
T_1 = 1 - \frac{\tilde Q}{r^4 + Q},
\ena

The nine-area is
\EQ
A_9 = \omega_5 L^4 (Q + \mu_1)^{3/4} (P + \mu_1)^{1/2},
\EN
which does not vanish in the ``extreme'' limit $\mu_1 \to 0$, as anticipated.
However, examining the condition of supersymmetry (\ref{ki}), one finds that
no supersymmetry is recovered in the limit $\mu_1 \to 0$. So it is true that
this is a solution of the field equations, but this limit simply implies that
the outer and inner horizons at $r=\mu_1^{1/4},0$ become degenerate.
In particular, it may not be easy to control quantum corrections even near
the limit.

The corresponding $(\td+3)$-dimensional Einstein-frame metric is
(here $\td=4$):
\EQ
ds_{\td+3}^2 = - \la^\td (r) f(r) dt^2 + \la^{-1}(r) \left[ f^{-1}(r) dr^2
 + r^2 d\Omega_{\td+1}^2 \right],
\label{metd}
\EN
\EQ
\la(r) = H^{-3/10} K^{-1/5}
 = \frac{r^2}{(r^4 + Q)^{3/10} (r^3 + P)^{1/5}}.
\label{exc1}
\EN

\subsection{$2^4 \subset 5$ with boost in $D=6$}

The boosted solution (\ref{stfi2}) with nonvanishing area has the metric
\bea
ds_{11}^2 &=& H^{2/3} \left[ H^{-5/3} (- K^{-1} f dt^2 + K \widehat{dy_1}^2 )
 + H^{-2/3}( dy_2^2 + \cdots + dy_5^2) \right. \nn
&& \left. + f^{-1} dr^2 + r^2 d\Omega_4^2 \right],
\label{15}
\ena
with $\td=3$. It does not seem possible to have
different charges in this case either. The solution is
\bea
&& F_{01\a r} = \frac{1}{2 \sqrt{3}} {T_1}',\hs{2} (\a=2,\ldots,5); \hs{2}
F_{abcd} = \shalf {T_2}', \nn
&& T_1 = 1 - \frac{\tilde Q}{r^3 + Q}; \hs{2}
T_2 = \frac{1}{\sqrt{3}}\left(1 - \frac{\tilde Q}{r^3}\right),
\ena

The nine-area is
\EQ
A_9 = \omega_4 L^5 (Q + \mu_1)^{5/6} (P + \mu_1)^{1/2},
\EN
which again does not vanish in the ``extreme'' limit but supersymmetry is
not recovered.

The corresponding six-dimensional Einstein-frame metric is (\ref{metd})
with $\td=3$ and
\EQ
\la(r) = H^{-5/12} K^{-1/4}
 = \frac{r^2}{(r^3 + Q)^{5/12} (r^3 + P)^{1/4}}.
\label{exc2}
\EN

\subsection{$2 \subset 5$ with boost in $D=6$}

The solution (\ref{25i}) with boost has the metric
\bea
ds_{11}^2 &=& H^{2/3} \left[ H^{-4/3}
 (- K^{-1} f dt^2 + K \widehat{dy_1}^2 + dy_2^2) + H^{-2/3}( dy_3^2
 + \cdots + dy_5^2) \right. \nn
&& \left. + f^{-1} dr^2 + r^2 d\Omega_4^2 \right],
\ena
with $\td=3$. The solution is
\bea
&& F_{012r} = \shalf {T_1}';\hs{2}
F_{abcd} = \shalf {T_2}', \nn
&& T_1= 1- \frac{\tilde Q}{r^3+Q}; \hs{2}
T_2 = \sqrt{\frac{2}{3}} \left(1- \frac{\tilde Q}{r^3}\right).
\ena

The nine-area is
\EQ
A_9 = \omega_4 L^5 \mu_1^{1/6} (Q + \mu_1)^{2/3} (P + \mu_1)^{1/2},
\EN
which vanishes in the ``extreme'' limit but supersymmetry remains broken.

The corresponding six-dimensional Einstein-frame metric is (\ref{metd})
with $\td=3$ and
\EQ
\la(r) = H^{-1/3} K^{-1/4}
 = \frac{r^{7/4}}{(r^3 + Q)^{1/3} (r^3 + P)^{1/4}}.
\label{exc3}
\EN

\subsection{Non-extreme two-brane with boost}

It is possible to discuss boosted versions of
$6\leq D(\equiv \td+3) \leq 9$ two-brane solutions (\ref{2}) together.
We take our metric as
\bea
ds_{11}^2 &=& H^{1/3} \left[ H^{-1} (- K^{-1} f dt^2 + K \widehat{dy_1}^2
 + dy_2^2 ) + dy_3^2 + \cdots + dy_{d-1}^2 \right. \nn
&& \left. + f^{-1} dr^2 + r^2 d\Omega_{\td+1}^2 \right].
\ena
We find that the solution to the field equation (\ref{fe}) is given
by~\cite{CT}
\bea
F_{012r} &=& \shalf T'; \hs{3}
T = 1 - \frac{\tilde Q}{r^\td + Q},
\label{nond}
\ena

The nine-area at the outer horizon $r=\mu_1^{1/\td}$ is given by
\bea
A_9 &=& \omega_{\td+1} L^{d-1} \left[ r^{\td+1} (K H)^{1/2}
 \right]_{r=\mu_1^{1/\td}} \nn
&=& \omega_{\td+1} L^{d-1} \mu_1^{1/\td} \sqrt{(Q+\mu_1)(P+\mu_1)}.
\label{aread}
\ena
In the extreme limit $\mu_1 \to 0$, this vanishes~\cite{CT,KT2}.
Here and in all the following examples, supersymmetry is recovered in the
extreme limit.

The corresponding $(\td+3)$-dimensional Einstein-frame metric is
(\ref{metd}) with
\EQ
\la(r) = (H K)^{-1/(\td+1)}
 = \frac{r^{2\td/(\td+1)}}{\left[ (r^\td + Q)(r^\td + P)\right]^{1/(\td+1)}}.
\label{la5}
\EN

\subsection{Non-extreme five-brane with boost}

The non-extreme generalization of the a five-brane (\ref{5}) with a boost
has the metrics
\bea
ds_{11}^2 = H^{2/3} \left[ H^{-1} (- K^{-1} f dt^2
 + K \widehat{dy_1}^2 + dy_2^2 + \cdots + dy_5^2) 
 + f^{-1} dr^2 + r^2 d\Omega_4^2 \right],
\label{5b}
\ena
with $\td=3$. We find 
\bea
F_{abcd} = \shalf T';\hs{3}
T = 1 - \frac{\tilde Q}{r^3}.
\ena
again solve the field equations~(\ref{fe})~\cite{CT}.

A similar analysis to the previous subsection shows that there is a regular
horizon at $r=\mu_1^{1/3}$ and that the nine-area vanishes in the extreme
limit. The expressions for nine-area and resulting five-dimensional metrics
are similar to (\ref{aread}) -- (\ref{la5}) with $\td=3$.

\subsection{Intersecting $2 \perp 2$ M-branes}

The final two-charge versions of higher-dimensional black hole solutions
we give explicitly are the intersecting $2\perp 2$ solutions (\ref{22}).
These exist in $D=6,7$ dimensions and we discuss them together.
The metric is
\bea
ds_{11}^2 &=& (H_1 H_2)^{1/3} \left[ - (H_1 H_2)^{-1} f dt^2
 + H_1^{-1} ( dy_1^2 + dy_2^2 ) + H_2^{-1} ( dy_3^2 + dy_4^2 )
 + dy_{d-1}^2 \right. \nn
&& \left. + f^{-1} dr^2 + r^2 d\Omega_{\td+1}^2 \right],
\ena
where $d=6$ or 5 ($D=\td+3=6$ or 7, respectively). The solution is
\bea
F_{012r} = \shalf {T_1}' ;\hs{3}
F_{034r} = \shalf {T_2}' ;\hs{3}
T_i = 1 - \frac{\tilde Q_i}{r^\td + Q_i}, (i=1,2),
\label{non22}
\ena
in an obvious notation.

The nine-area at the outer horizon $r=\mu_1^{1/\td}$ is given by
\bea
A_9 &=& \omega_{\td+1} L^{d-1} \left[ r^{\td+1} (H_1 H_2)^{1/2}
 \right]_{r=\mu_1^{1/\td}} \nn
&=& \omega_{\td+1} L^{d-1} \mu_1^{1/\td} \sqrt{(Q_1+\mu_1)(Q_2+\mu_1)},
\label{area22}
\ena
which vanishes in the extreme limit $\mu_1 \to 0$.
The corresponding $(\td+3)$-dimensional Einstein-frame metric is
(\ref{metd}) with $\la(r)$ almost the same as (\ref{la5}).

\sect{ADM mass and entropy}

Upon dimensional reduction of the world-volumes ($(d-1)$-dimensional $y_\a$
space) of the intersecting M-branes, we obtain the black hole solutions in
$\td +3$ dimensions. From eqs.~(\ref{4met}), (\ref{la1}), (\ref{la2}),
(\ref{5met}), (\ref{la3}), (\ref{la4}) and (\ref{metd}), we see that, except
for the special cases in (\ref{exc1}), (\ref{exc2}) and (\ref{exc3}),
the Einstein-frame metrics can be written universally as
\EQ
ds_{\td+3}^2 = - \la^\td (r) f(r) dt^2 + \la^{-1}(r) \left[ f^{-1}(r) dr^2
 + r^2 d\Omega_{\td+1}^2 \right],
\EN
where
\bea
\la(r) = (H_1 H_2 \cdots H_n)^{-1/(\td+1)}; \hs{2}
H_i = 1 + \frac{Q_i}{r^\td}, \hs{3} (i=1,\ldots ,n),
\ena
$f$ is given in (\ref{nondf}) and we have expressed all charges by $Q_i$.
For $\td \geq 3$, we should put $\mu_2=0$.

From the asymptotic form of the metric $g_{00}$, we can read off the ADM
mass~\cite{MP}:
\EQ
M_{ADM} = a \left[(\td + 1) (\mu_1 + \mu_2) + \td \sum_{i=1}^n Q_i \right],
\label{adm}
\EN
where the constant $a$ is defined by
\EQ
a = \frac{\omega_{\td+1}}{2 \kappa^2} L^{d-1},
\EN
and we have defined the eleven-dimensional Newton's constant as
$G_{11}= \kappa^2/8\pi$. For the solutions (\ref{exc1}) $(\td=4)$ and
(\ref{exc2}) $(\td=3)$,
\EQ
M_{ADM}^{exc} = a \left[ (\td+1) \mu_1 + (\td+2)Q + \td P \right],
\label{adme}
\EN
For (\ref{exc3}), we find
\EQ
{M_{ADM}^{exc}} = a \left[ 4 \mu_1 + 4 Q + 3 P \right],
\label{adme1}
\EN

If we use, instead of $Q_i$, the charges
\EQ
P_i \equiv \frac{\mu_1-\mu_2}{2} \sinh 2\d_i,
\EN
which are also fixed in the extreme limit, the ADM mass~(\ref{adm})
is cast into the form
\EQ
M_{ADM} = a \td \left[ \sum_{i=1}^n \sqrt{ P_i^2
 + \left( \frac{\mu_1 - \mu_2}{2} \right)^2} + \la ( \mu_1 + \mu_2 ) \right],
\label{adm1}
\EN
where the constant $\la$ is defined by
\EQ
\la \equiv \frac{\td+1}{\td} - \frac{n}{2}.
\EN
This constant $\la$ vanishes in our solutions for $D=4,(n=4)$ and $D=5,(n=3)$
in sects.~3, 4; $n=1$ and $\la=\frac{\td+2}{2 \td}$ for two- and five-branes
without boost in sect.~6; for other usual solutions in higher dimensions,
$n=2$ and $\la=\frac{1}{\td}$. If we try to write (\ref{adme}) as
(\ref{adm1}), it takes the form
\EQ
M_{ADM}^{exc} = a\left[ \td\sqrt{ P_1^2 + \left( \frac{\mu_1}{2} \right)^2}
 + (\td+2) \sqrt{ P_2^2 + \left( \frac{\mu_1}{2} \right)^2} \right],
\label{adm1e}
\EN
in an obvious notation and an ``effective'' constant $\la$ is zero.
Similarly, for (\ref{adme1}), one finds that the effective
$\la=\frac{1}{6}$.

The constant $\la$ gives the measure how the area or entropy vanish in the
extreme limit: If it is positive, they vanish like $\sim \mu_1^\la$; if
it is zero, they remain finite.

The entropy is given by
\bea
S_{BH} &=& \frac{2\pi A_9}{\kappa^2}
 = 4\pi a \mu_1^\la \prod_{i=1}^n \left[ (\mu_1 - \mu_2)^{1/2} \cosh \d_i
 \right] \nn
&=& 4\pi a (\mu_1 - \mu_2)^{\frac{\td+1}{\td}-\la} \mu_1^\la
 \prod_{i=1}^n \cosh \d_i.
\label{ent}
\ena
This agrees with the similar expression in ref.~\cite{CT}.
Again for (\ref{14}) and (\ref{15}), one has
\bea
S_{BH}^{exc} = \frac{2\pi A_9}{\kappa^2}
 = 4\pi a \mu_1^{(\td+1)/\td} (\cosh \d_1)^{(\td+2)/\td} \cosh \d_2.
\label{ente}
\ena

Let us next compute the Hawking temperature from the periodicity of the
Euclideanized geometry. Suppose that our metric is of the form
\EQ
ds^2 = e^{2 u} d\tau^2 + e^{2 v} dr^2 + r^2 d\Omega_{\td+1}^2,
\label{nmet}
\EN
and $e^{2u}=0$ at the horizon $r=r_0$. Then near the horizon, the metric
(\ref{nmet}) can be written as
\EQ
ds^2 = \left[ \left( e^{u} \right)' e^{-v}|_{r=r_0} \right]^2 R^2 d\tau^2
 + dR^2 + r^2 d\Omega_{\td+1}^2,
\EN
where $R=0$ at $r=r_0$ and $r$ should be regarded as a function of $R$.
We see that $\tau$ can be interpreted as angular
coordinate on the $R-\tau$ plane.  From the condition that there is no
singularity at $R=0$, we find that $\tau$ should be a periodic variable
with period $\b=1/T_H$ with
\EQ
2 \pi T_H = (e^u)' e^{-v} |_{r=r_0}.
\EN
This formula gives the Hawking temperature for our normal solutions
\EQ
T_H = \frac{\td}{4 \pi} (\mu_1 - \mu_2)^{\la-1/\td} \mu_1^{-\la}
 \prod_{i=1}^n (\cosh \d_i)^{-1},
\EN
and the entropy (\ref{ent}) can be written as
\EQ
S_{BH} = a \td \frac{\mu_1 - \mu_2}{T_H}.
\EN
Expressed in terms of the charges and $\mu_1,\mu_2$, the entropy
can be transformed into
\bea
S_{BH} &=& 4 \pi a \mu_1^\la \prod_{i=1}^n \sqrt{Q_i + \mu_1} \nn
 &=& 4 \pi a \mu_1^\la \prod_{i=1}^n \left[
 \sqrt{P_i^2 + \left( \frac{\mu_1 - \mu_2}{2} \right)^2}
 + \frac{\mu_1 - \mu_2}{2} \right]^{1/2}.
\label{ent2}
\ena
For the exceptional cases, the powers of the charges in the product change,
but the power of $\mu_1$ in front remains the same.

The extreme limit is to send $\mu_1 \to \mu_2 (=0$ for $D\geq 6)$ with
$(\mu_1-\mu_2)^{1/2} \cosh\d_i$ fixed. In this limit, the entropy
(\ref{ent}) for $\la=0$ and (\ref{ente}) is nonvanishing but
the Hawking temperature vanishes in our all solutions.

Finally we note that the ADM mass $M_{ADM}$ in our solutions in
$D=4, 5$ dimensions in sects.~3 and 4 resembles the energy of a system
of relativistic particles with masses $P_i$ and momenta proportional to
$(\mu_1-\mu_2)$. For other higher dimensional solutions, the ADM mass does
not have such an interpretation.

\sect{Conclusions}

We have presented non-extreme solutions in the form with two nonvanishing
deformation parameters, similar to the four-dimensional Reissner-Nordstr{\o}m
black holes. They turn out to be equivalent to those extensively studied in
ref.~\cite{CT} under the replacement~(\ref{repl}) for $D=4$ and 5 dimensions.

We have also tried to find higher-dimensional solutions. We have carried out
a rather systematic analysis of the solutions in the case of equal charges
and reduced the problem to a simple algebraic one. In the process, we were
able to show that the generalization of these solutions with {\it two
nonvanishing} deformation parameters is impossible, and that the area of
the black holes in general vanishes in the extreme limit for static
spherically-symmetric ones in higher dimensions. The only exceptions are
the solutions (\ref{exc1}) in $D=7$ and (\ref{exc2}) in $D=6$ dimensions.
Unfortunately supersymmetry remains broken in the limit even though the
inner and outer horizons coincide and the Hawking temperature vanishes.
If we insist that supersymmetry
should be recovered in the extreme limit, these exceptional solutions
may not be regarded as ``good'' solutions. Together with the results
in refs.~\cite{CYn,CT}, our results imply that the extreme solutions
in $D \geq 6$ necessarily involve naked singularities or do not have
regular horizon of finite area except for (\ref{exc1}) and (\ref{exc2})
as long as we consider static spherically-symmetric solutions.

It would be quite interesting to give a statistical explanation of the entropy
[8-16,21,22,28,29] and examine the Hawking radiation for these solutions,
in order to gain insight into the quantum M-theory and also to understand
what new features emerge from this kind of investigation.
Our algebraic results should also be useful for finding new solutions.
We hope to discuss these questions elsewhere.

\section*{Acknowledgements}

We would like to thank R. Emparan, I. Klebanov and A. Tseytlin for pointing out
the relations of our solutions in $D=4,5$ to those in refs.~\cite{T,KT2,CT}
by the redefinition~(\ref{repl}). Special thanks are due to A. Tseytlin for
useful correspondence which helped us to improve the manuscript.

\newpage
\newcommand{\NP}[1]{Nucl.\ Phys.\ {\bf #1}}
\newcommand{\AP}[1]{Ann.\ Phys.\ {\bf #1}}
\newcommand{\PL}[1]{Phys.\ Lett.\ {\bf #1}}
\newcommand{\NC}[1]{Nuovo Cimento {\bf #1}}
\newcommand{\CMP}[1]{Comm.\ Math.\ Phys.\ {\bf #1}}
\newcommand{\PR}[1]{Phys.\ Rev.\ {\bf #1}}
\newcommand{\PRE}[1]{Phys.\ Rep.\ {\bf #1}}
\newcommand{\PRL}[1]{Phys.\ Rev.\ Lett.\ {\bf #1}}
\newcommand{\PTP}[1]{Prog.\ Theor.\ Phys.\ {\bf #1}}
\newcommand{\PTPS}[1]{Prog.\ Theor.\ Phys.\ Suppl.\ {\bf #1}}
\newcommand{\MPL}[1]{Mod.\ Phys.\ Lett.\ {\bf #1}}
\newcommand{\IJMP}[1]{Int.\ Jour.\ Mod.\ Phys.\ {\bf #1}}
\newcommand{\JP}[1]{Jour.\ Phys.\ {\bf #1}}


\begin{thebibliography}{99}
\bibitem{DGH} A. Dabholkar, G. Gibbons, J. A. Harvey and F. Ruiz Ruiz,
 \NP{B340} (1990) 33; for a review and earlier references, see M. Duff,
 R. Khuri and J. X. Lu, \PRE{259} (1995) 213, hep-th/9412184.
\bibitem{HS} G. T. Horowitz and A. Strominger, \NP{B360} (1991) 197.
\bibitem{GU} R. G\"uven, \PL{B276} (1992) 49.
\bibitem{Sen} A. Sen, \NP{B440} (1995) 421, hep-th/9411187; \\
 G. T. Horowitz and A. Sen, \PR{D53} (1996) 808, hep-th/9509108.
\bibitem{CYe} M. Cveti\v c and D. Youm, \PR{D53} (1996) R584,
 hep-th/9507090; preprint, hep-th/9512127; \NP{B476} (1996) 118,
 hep-th/9603100.
\bibitem{CYn} M. Cveti\v c and D. Youm, Proceedings of String '95,
 hep-th/9508058;\PR{D54} (1996) 2612, hep-th/9603147; \NP{B477} (1996)
 118, hep-th/9605051; preprint, hep-th/9612229.
\bibitem{CT1} M. Cveti\v c and A. A. Tseytlin, \PR{D53} (1996) 5619,
 hep-th/ 9512031;\\
 A. A. Tseytlin, \MPL{A11} (1996) 689, hep-th/9601177.
\bibitem{SV} A. Strominger and C. Vafa, \PL{B379} (1996) 99, hep-th/9601029;
 A. Strominger, \PL{B383} (1996) 39, hep-th/9602111.
\bibitem{CM} C. G. Callan and J. M. Mardacena, \NP{B472} (1996) 591,
 hep-th/9602043.
\bibitem{BMPV} J. C. Breckenridge, R. C. Myers, A. W. Peet and C. Vafa,
 preprint, hep-th/9602065;\\
 J. C. Breckenridge, D. A. Lowe, R. C. Myers, A. W. Peet, A. Strominger
 and C. Vafa, \PL{B381} (1996) 423, hep-th/9603078.
\bibitem{MSS} J. Maldacena and L. Susskind, \NP{B475} (1996) 679,
 hep-th/9604042.
\bibitem{MS} J. M. Maldacena and A. Strominger, \PRL{77} (1996),
 hep-th/9603060.
\bibitem{JKM} C. V. Johnson, R. R. Khuri and R. C. Myers, \PL{B378} (1996)
 78, hep-th/9603061.
\bibitem{HS1} G. Horowitz and A. Strominger, \PRL{77} (1996) 2368,
 hep-th/9602051.
\bibitem{HMS} G. T. Horowitz, J. M. Maldacena and A. Strominger,
 \PL{B383} (1996) 151, hep-th/9603109.
\bibitem{HLM} G. T. Horowitz, D. A. Lowe and J. M. Maldacena, \PRL{77}
 (1996) 430, hep-th/9603195.
\bibitem{PT} G. Papadopoulos and P. Townsend, \PL{B380} (1996) 273,
 hep-th/9603087.
\bibitem{T} A. A. Tseytlin, \NP{B475} (1996)149, hep-th/9604035.
\bibitem{DLP} M. J. Duff, H. L\"u and C. N. Pope, \PL{B382}
 (1996) 73, hep-th/9604052.
\bibitem{PO} H. L\"u, C. N. Pope, E. Sezgin and K. S. Stelle, \NP{B456}
 (1996) 669, hep-th/9508042;
 N. Khviengia, Z. Khviengia, H. L\"u and C. N. Pope, \PL{B388}
 (1996) 21, hep-th/9605077.
\bibitem{KT1} I. R. Klebanov and A. A. Tseytlin, \NP{B475} (1996) 164,
 hep-th/9604089.
\bibitem{KT2} I. R. Klebanov and A. A. Tseytlin, \NP{B475} (1996) 179,
 hep-th/9604166.
\bibitem{CT} M. Cveti\v c and A. A. Tseytlin, \NP{B478} (1996) 181,
 hep-th/ 9606033.
\bibitem{GA} J. P. Gauntlett, D. A. Kastor and J. Traschen, \NP{B478} (1996)
 544, hep-th/9604179.
\bibitem{DR} M. J. Duff and J. Rahmfeld, \NP{B481} (1996) 332, hep-th/9605085.
\bibitem{CH} M. Cveti\v c and C. M. Hull, \NP{B480} (1996) 296, hep-th/9606193.
\bibitem{GAR} D. Garfinkle, \PR{D46} (1992) 4286.
\bibitem{MP} R. C. Myers and M. J. Perry, \AP{172} (1986) 304.
\bibitem{BL} V. Balasubramanian and F. Larsen, \NP{B478} (1996) 199,
 hep-th/9604189.
\bibitem{BE} K. Behrndt, E. Bergshoeff, \PL{B383} (1996) 383, hep-th/9605216.
\bibitem{CO} M. S. Costa, preprints, hep-th/9609181, hep-th/9610138.
\end{thebibliography}
\end{document}